PLoS one

# Local Perturbations Do Not Affect Stability of Laboratory Fruitfly Metapopulations

Sutirth Dey, Amitabh Joshi*

Evolutionary Biology Laboratory, Evolutionary and Organismal Biology Unit, Jawaharlal Nehru Centre for Advanced Scientific Research, Bangalore, India

*Background*. A large number of theoretical studies predict that the dynamics of spatially structured populations (metapopulations) can be altered by constant perturbations to local population size. However, these studies presume large metapopulations inhabiting noise-free, zero-extinction environments, and their predictions have never been empirically verified. *Methodology/Principal Findings*. Here we report an empirical study on the effects of localized perturbations on global dynamics and stability, using fruitfly metapopulations in the laboratory. We find that constant addition of individuals to a particular subpopulation in every generation stabilizes that subpopulation locally, but does not have any detectable effect on the dynamics and stability of the metapopulation. Simulations of our experimental system using a simple but widely applicable model of population dynamics were able to recover the empirical findings, indicating the generality of our results. We then simulated the possible consequences of perturbing more subpopulations, increasing the strength of perturbations, and varying the rate of migration, but found that none of these conditions were expected to alter the outcomes of our experiments. Finally, we show that our main results are robust to the presence of local extinctions in the metapopulation. *Conclusions/Significance*. Our study shows that localized perturbations are unlikely to affect the dynamics of real metapopulations, a finding that has cautionary implications for ecologists and conservation biologists faced with the problem of stabilizing unstable metapopulations in nature.

## INTRODUCTION

Simple one-dimensional maps can exhibit a variety of dynamic behaviors ranging from stable points to limit cycles to chaos [1,2], and have been extensively used to model the dynamics of single populations. It has been shown that for large ranges of parameter values, the dynamics of such maps can be substantially altered by the addition [3] or removal [4,5] of a constant number of individuals every generation. This happens because such perturbations can change the slope of the return map at the equilibrium point, thereby affecting the dynamics of the population [6,7]. However, such simple models explicitly assume that the individuals in the population are homogeneously distributed in space, whereas many real populations exhibit spatial structuring into metapopulations: groups of local populations (subpopulations) connected by migration. Many methods for stabilizing the dynamics of metapopulations by perturbation have been proposed in the context of both ecological [8–10] as well as physical [11–14] systems, and some of these proposed algorithms have been empirically verified in physical [15,16] or in-vitro physiological systems [17,18]. However, to the best of our knowledge, there has been no experimental confirmation of stabilization of a real biological metapopulation by constant perturbation.

There are several reasons why experimental studies lag far behind the substantial body of theoretical predictions on the issue of metapopulation stabilization by perturbation. Most theoretical studies on the subject have explicitly concentrated on stability in terms of amelioration of chaos to get stable points or limit cycles [8–10,19]. However, since real organisms come in discrete (integer) numbers, no real population can exhibit chaos in the strict sense, although this does not rule out the possibility of complex dynamics [20]. Moreover, most theoretical treatments assume a large number of subpopulations in an ideal, noise free, zero-extinction environment, which is far from the reality of actual biological metapopulations.

Here, we investigate whether pinning can stabilize the dynamics of real metapopulations that are generally noisy, finite (often small)-sized and prone to local extinctions. We begin by reporting a 21-generation long experiment on the effects of localized perturbations at the subpopulation level on local and global stability, using two sets of four replicate *Drosophila melanogaster* metapopulations each. Each metapopulation contained 9 subpopulations, represented by single vial cultures, arranged on the periphery of a circle, with 30% migration in each generation to the two nearest neighbors. In the four *pinned* [10] metapopulations, we perturbed the same subpopulation (henceforth, the *pinned subpopulation*) in every generation by adding a fixed number of flies from outside the system, whereas there were no such perturbations in the four *control* metapopulations. We show that although pinning affects the dynamics of the particular pinned subpopulation, it has no measurable effects on metapopulation dynamics. We also show that Ricker-based simulations capture the patterns observed in the data, indicating that our results are generalizable. We further demonstrate, via simulations, that our findings are robust to the various assumptions made in the

. . . . . . . . . . . . . . . . . . . . . . . . . . . . . . . . . . . . . . . . . . . . . . .

Funding: This work was supported by funds from the Department of Science and Technology, Government of India, to A. Joshi. S. Dey was supported by a Senior Research Fellowship from the Council for Scientific and Industrial Research, Government of India.

Competing Interests: The authors have declared that no competing interests exist.

* To whom correspondence should be addressed. E-mail: ajoshi@jncasr.ac.in

experiment regarding the number of pinned patches per metapopulation, the strength of pinning, and migration rates. Finally, we investigate the effects of the interaction of extinction and pinning in shaping metapopulation dynamics and show that our results generally hold even in the absence of local extinctions. Since we explicitly focus on indicators of stability that are ecologically meaningful and can be measured relatively easily, our results are not only of interest to ecologists but have potential practical implications for conservation biologists trying to develop schemes for stabilizing a fragmented population.

## RESULTS AND DISCUSSION

### Experiment

We found that the mean fluctuation index, FI [21], of the pinned subpopulations was significantly lower ($F_{1,3} = 180.95$, $p < 0.0009$) than the mean FI of the remaining eight subpopulations in the pinned metapopulations (Fig 1A). This indicates that constant addition of flies every generation from outside the metapopulation stabilized the pinned subpopulation by reducing the fluctuation in its population size from one generation to the next. We then sought to check whether this stabilized subpopulation (i.e. the pinned subpopulation) was in turn able to affect the dynamics in its neighborhood. For this, we divided the pinned metapopulation into three groups; each consisting of three subpopulations. The *pinned* group contained the pinned subpopulation and its two immediate neighbors, while the other two groups (*No pin 1* and *No Pin 2*) comprised of the three neighboring subpopulations to the right and left of the pinned group, respectively. There was no significant difference ($F_{2,6} = .64$, $p < 0.56$) between the average FI of the pinned group and the neighboring groups (Fig 1B), thus indicating that the reduced FI of the pinned subpopulation does not translate into significant stabilization of the pinned group vis-à-vis the neighboring non-pinned groups.

We then measured various attributes of metapopulation dynamics (see Materials and methods), like metapopulation stability (Fig 2A), subpopulation stability (Fig 2B), synchrony among nearest neighboring subpopulations (Fig 2C), and average subpopulation size (Fig 2D), but did not observe any significant difference between the control and the pinned metapopulations. Another commonly used measure of population stability, namely the coefficient of variation (CV) of population size, was also found to be similar in both treatments at the metapopulation ($F_{1,6} = .32$, $p < 0.59$) and subpopulation ($F_{1,6} = 1.13$, $p < 0.33$) level. When we defined an extinct patch as one that remained empty during breeding after migration had taken place, the total number of subpopulation extinctions over 21 generations was considerably less in the pinned metapopulations (39) than in the controls (69). However, this is an artifact of the experimental protocol, as all three subpopulations in the pinned group of the pinned metapopulation were, by design, receiving flies from outside every generation (see Materials and methods) and hence they were never scored as extinct. When we considered pre-migration extinction, in the form of absence of at least one breeding pair (i.e. 1 male+1 female) in a subpopulation, there was no difference in number of extinctions per generation between the pinned and control metapopulations ($F_{1,6} = .009$, $p < 0.93$), indicating that pinning did not affect the persistence of subpopulations. Together, these observations suggest that pinning had no effects on stability or any of the other measured attributes of the metapopulation.

The above experimental observations could have arisen due to two possible reasons: either pinning, at least at the levels used here, genuinely does not affect the dynamics of metapopulations, or there are some unique features of *Drosophila* life-history or ecology in the laboratory that ameliorate the effects of pinning. In case the second hypothesis were true, these results are not likely to be generalizable to other species, and hence would be of limited interest. One way to distinguish between these competing hypotheses is to simulate our experimental system with a biologically relevant model of population dynamics that is broadly applicable to several species and does not include any specific features of *Drosophila* life-history or laboratory ecology. If such a model were able to capture at least the general features seen in the experimental data, then one would expect our results to be valid for a wide spectrum of organisms.

### Simulations

**Experimental system** It has been analytically demonstrated that populations with a random spatial distribution and scramble competition follow the Ricker dynamics [22]. Since laboratory cultures of *Drosophila* exhibit both features, we modeled subpopulation dynamics by the Ricker model [23], a simple one-dimensional model of population dynamics, whose qualitative behavior is solely determined by the intrinsic growth rate parameter, $r$ [24]. The Ricker model has been shown to be a good descriptor of the dynamics of various types of organisms including microbes [25], fishes [26] and insects [27], including *Drosophila* [21,28]. Thus, this model satisfies the criteria of being biologically relevant, non-*Drosophila* specific, and widely applicable.

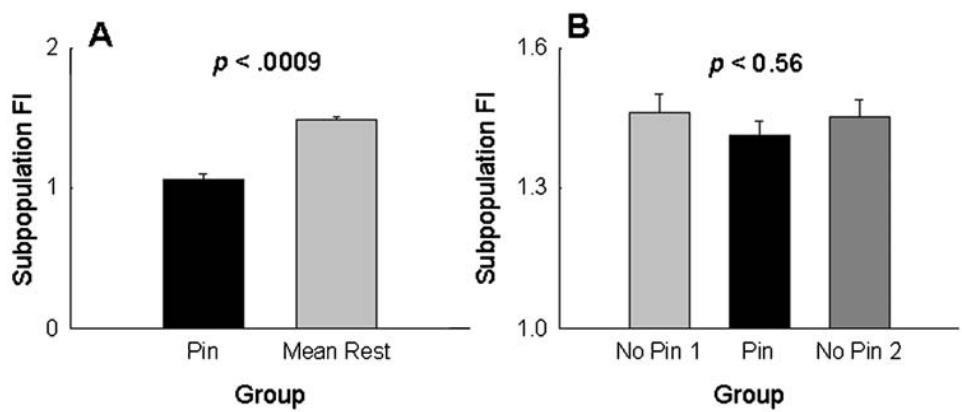

**Figure 1.** Experiment: effect of pinning at the subpopulation level, averaged over four replicate metapopulations. (A) The mean FI of the pinned subpopulation was significantly less than the mean of the remaining eight subpopulations. (B) There was no difference in the average FI of the pinned group (the pinned subpopulation and its two immediate neighbors) and the two neighboring groups on either side (No Pin1 and No Pin 2). This suggests that the stabilized subpopulation could not stabilize the dynamics of the pinned group vis-à-vis the two neighboring groups. Error bars indicate standard errors around the mean in this and all subsequent figures.
doi:10.1371/journal.pone.0000233.g001

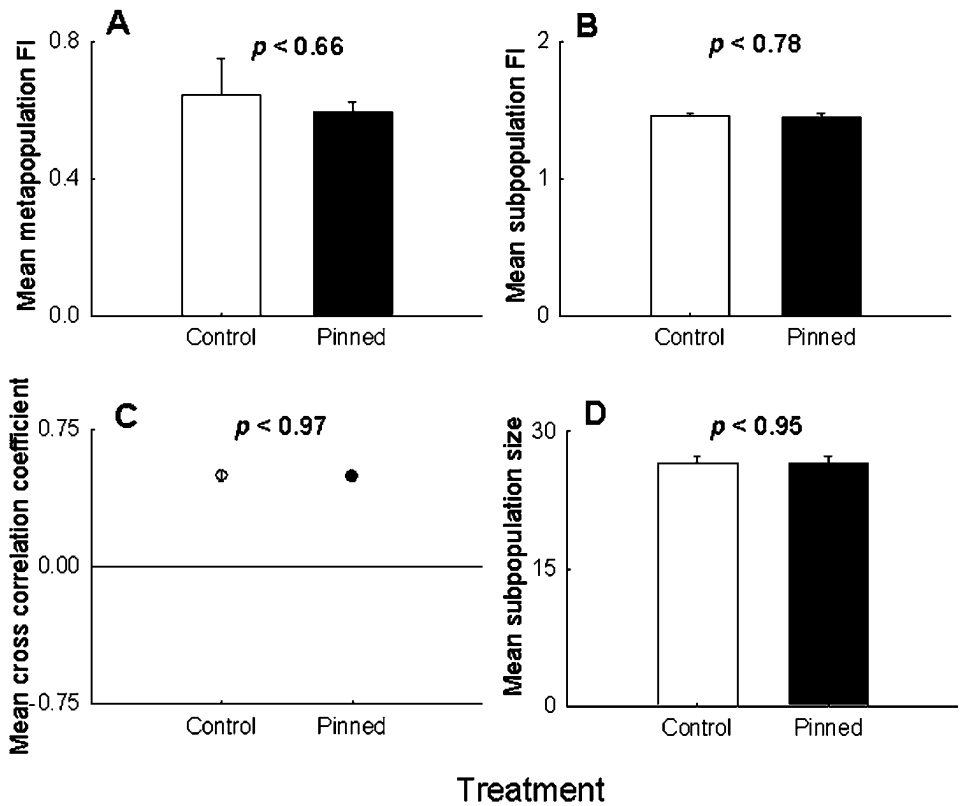

**Figure 2.** Experiment: effect of pinning at the metapopulation level, averaged over four replicate metapopulations. (A) Metapopulation stability and (B) subpopulation stability were measured as the fluctuation index (FI) over 21 generations. (C) Synchrony among nearest neighbors was measured as the cross-correlation at lag zero of the first differenced ln-transformed values of population sizes. Due to the high rates of migration, the subpopulations were found to be in synchrony, as demonstrated by the positive cross correlation coefficients. (D) Average subpopulation size. There was no difference among the pinned and the control metapopulations in any of the panels, indicating that pinning had no detectable effect on metapopulation dynamics.
doi:10.1371/journal.pone.0000233.g002

The simulation results were seen to support the experimental observations. The FI of the pinned subpopulation in the simulations was found to be lower than the mean of the remaining eight subpopulations for a range of $r$ values (Fig 3A), while the mean FI of the pinned group was found to be similar, or - for some values of $r$ - slightly lower than the other groups (Fig 3B). As in the experiments, there were no observable differences in the metapopulation FI (Fig 4A), subpopulation FI (Fig 4B) or subpopulation synchrony (Fig 4C) between control and pinned metapopulations in the simulations. The model predicted a slight decrease in subpopulation size (Fig 4D) in the pinned metapopulations, at least for lower values of intrinsic growth rate $r$, an effect that was not observed in the experiments. Overall, these simulation results agreed well with the experimental data, suggesting that our observations are unlikely to be *Drosophila*-specific. It is important to note here that these results do not invalidate previous theoretical studies on using regular perturbations to stabilize chaotic systems to get limit cycles or stable points [9,10], as those studies investigated a different kind of stability altogether. Our findings merely suggest that, all else being equal, the effects of localized perturbations are unlikely to be measurable at the metapopulation level in real biological populations.

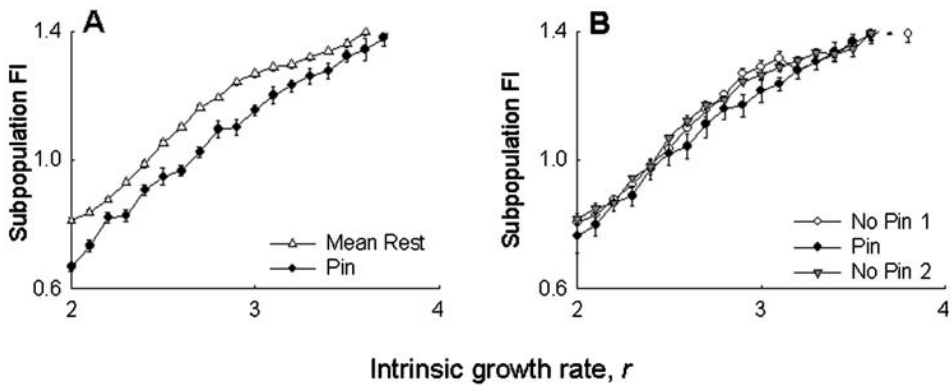

**Figure 3.** Simulations mimicking experiment: effect of pinning at the subpopulation level. (A) The FI of the pinned subpopulation was lower than the mean of the remaining eight subpopulations, over a substantial range of the intrinsic growth rate parameter, $r$. (B) There was no difference in the average FI of the pinned group and its two neighboring groups. These observations agree with the experimental results (cf. Fig 1), implying that the experimental findings are non-*Drosophila* specific. All data points in this, and subsequent simulation figures, represent average of 10 independent runs. See text for details of simulations.
doi:10.1371/journal.pone.0000233.g003

**Relaxing experimental assumptions** Studies on laboratory systems generally entail a higher degree of accuracy in measurement and better control over noise than is possible in nature. Thus, failure to observe an effect of pinning under controlled laboratory conditions indicates that, at least under conditions similar to the experiment, pinning is expected to be of limited importance in controlling the dynamics of real populations. However, earlier theoretical studies have shown that the number of patches pinned, the magnitude of pinning, and the migration rate can affect the dynamics of the metapopulation [9,10,19,29]. Since we conducted the experiments under a fixed set of conditions - pinning one patch with 8 females in each generation, under 30% migration rates - it is natural to ask whether our results would have been altered if one or more of these conditions had been different. Moreover, in this study, we used unstable *Drosophila* subpopulations that had a high rate of extinction, which too can possibly influence the dynamics. Although the ideal way to address these issues would have been to conduct more experiments under appropriate conditions, logistic constraints prevented us from doing so. Since this and earlier studies [21] have indicated that Ricker-based coupled map lattices are good surrogates for *Drosophila* metapopulations, we used the same simulation framework described above to investigate the effects of departures from the experimental conditions.

Increasing the proportion of pinned patches did not change the metapopulation FI, at least for values of $r<3$ (Fig 5A). For higher values of $r$, which signifies the chaotic zone in case of the Ricker model, increasing the number of pinned patches generally increased the metapopulation FI (hence instability), although there were no distinct patterns (Fig 5A). Altering the pinning strength failed to produce any observable change in the metapopulation dynamics (Fig 5B). It has been shown earlier that low and high rates of migration reduce and enhance the metapopulation FI, respectively [21]. While similar patterns were observed in our simulations, there was no observable difference between the control (Fig 6A) and the pinned (6B) metapopulations. Together, these observations suggest that our experimental results are robust and no qualitative changes in the outcome would have been expected, even if the experiments were conducted under different conditions of pinning or migration rates.

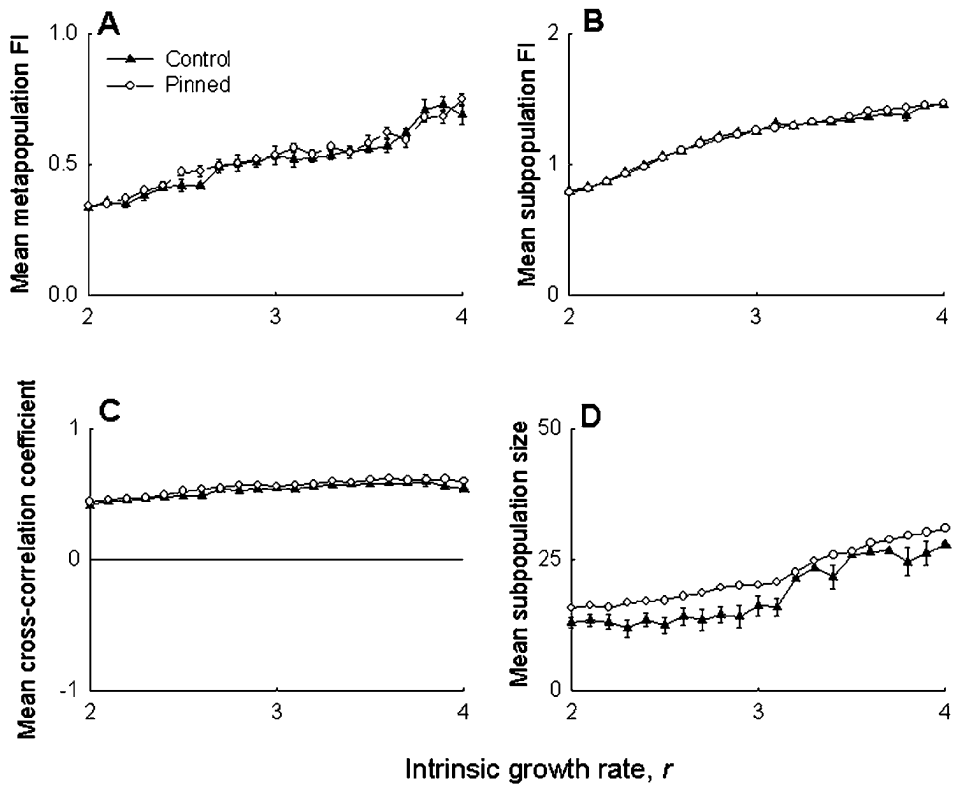

**Figure 4.** Simulations mimicking experiment: effect of pinning at the metapopulation level. Ricker based simulations predicted no difference in (A) metapopulation stability, (B) subpopulation stability, and (C) synchrony amongst nearest-neighbors, between the control and pinned metapopulations. (D) The simulations suggested a slight decrease in subpopulation size for low values of $r$, which was not picked up by the experiment. Overall, these results agree with the corresponding experimental findings (Fig 2), indicating that they are likely to be applicable to other species.
doi:10.1371/journal.pone.0000233.g004

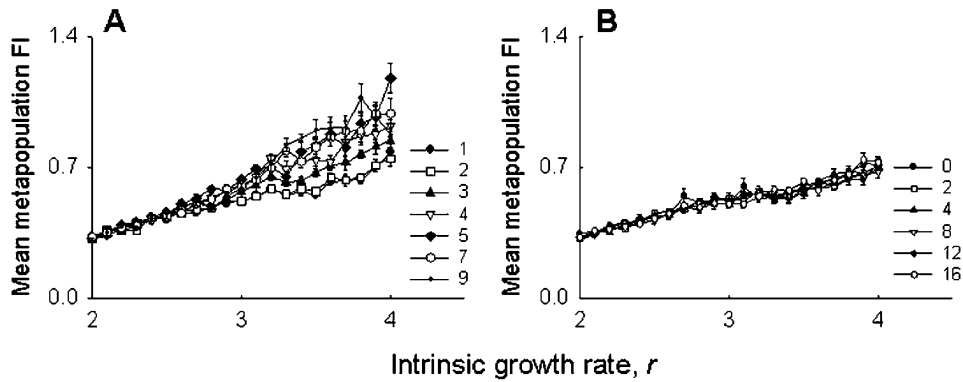

**Figure 5. Simulations relaxing experimental assumptions: effect of pinning density and magnitude on stability.** (A) There was no effect on metapopulation FI due to pinning greater number of patches for $r<3$. When $r>3$, increasing the proportion of pinned patches generally increased FI, although there were no consistent patterns. (B) Varying the magnitude of pinning had no effects on metapopulation stability. These suggest that the empirical results are robust to departures from the conditions of the experiment.
doi:10.1371/journal.pone.0000233.g005

**Absence of extinction** Like most of their natural counterparts, our experimental metapopulations experienced frequent local extinctions followed by recolonization from neighboring patches. This raises the question as to whether the observed effects of pinning were modulated by subpopulation extinctions. We investigated this issue by repeating all the above-mentioned simulations in the absence of any extinctions. When there were no extinctions, the FI of the pinned subpopulation was found to be slightly less than the mean of the remaining eight subpopulations (Fig 7A) for $r<2.6$. However, for $r>2.6$, the pinned subpopulation had a higher FI, which agrees with the findings of a previous study that used an individual based model without any local extinction [30] but is contrary to our experimental data (Fig 1A) and our earlier simulation (Fig 3A). The FI of the pinned group was also higher than the two neighboring groups (cf Fig 7B and Fig 3B) for $r>2.6$. These observations indicate that in the absence of extinction, the effect of pinning on subpopulation dynamics interacts with the intrinsic growth rate of the subpopulations. However, in the presence of local extinctions, pinning appears to uniformly stabilize the subpopulation dynamics.

These differences at the subpopulation level, however, did not lead to major changes in the results at the metapopulation level

(Fig 8A) compared to the case when extinction probabilities were explicitly incorporated into the simulations (Fig 4A). Thus, although there seemed to be an effect of pinning on the shape of the metapopulation FI profile (cf Fig 8A and Fig 4A), there were no systematic differences in the FI of the control and the pinned metapopulations. The subpopulation FI (Fig 8B) and the nearest neighbor cross-correlation coefficient (Fig 8C) were also seen to be similar in controlled and pinned metapopulations. Under no extinction, the model predicted an increase in average subpopulation size of the pinned metapopulations (Fig 8D) for high values of $r$ ($>3.2$), which was again different from the effects under extinction (Fig 4D). Together, these observations suggest that while the subpopulation level dynamics under pinning might be affected by the presence/absence of extinction, this difference is unlikely to have a major global impact at the metapopulation level.

When there were no extinctions, increasing the number of pinned subpopulations had no effects at low values of $r$ but, in general, destabilized the metapopulations by increasing the FI for high values of $r$ (Fig 9A). Although there was a distinct change in the profile, and an increase in the overall magnitude of metapopulation FI (cf Fig 9A and Fig 5A), the basic observation that increasing the proportion of pinned subpopulations generally increased the metapopulation FI, remained unchanged. The prediction that increasing the density of pinned subpopulations might lead to an observable change in the global dynamics, at least for a sizable range of $r$-values, agrees well with previous results

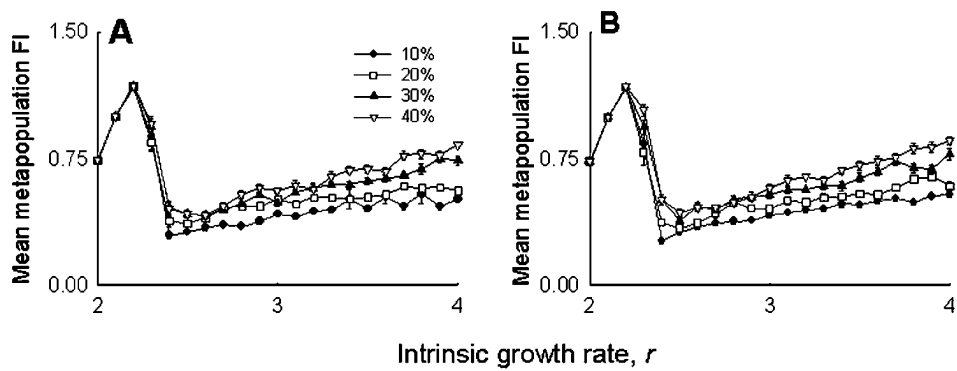

**Figure 6. Simulations relaxing experimental assumptions: effect of migration rate on stability.** Various rates of migration did not have a differential effect on the stability of the (A) control and (B) pinned metapopulation, again indicating the robustness of the experimental findings.
doi:10.1371/journal.pone.0000233.g006

  

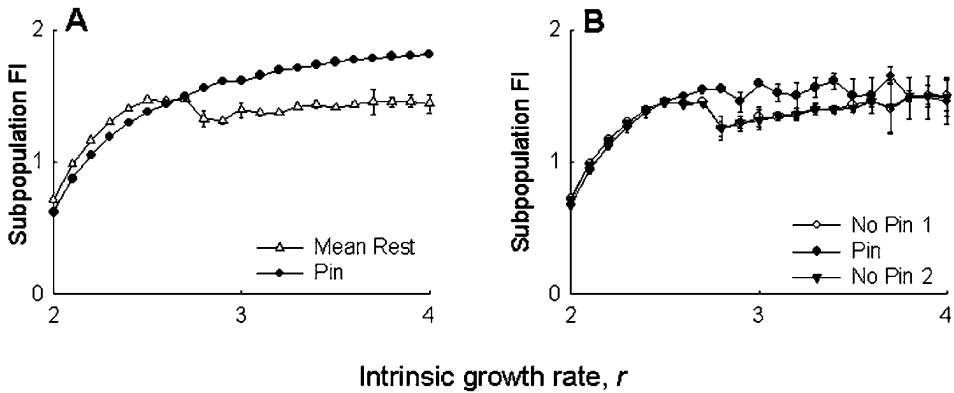

**Figure 7. Simulations with no extinctions: effect of pinning at the subpopulation level.** (A) The FI of the pinned subpopulation was higher than the mean of the remaining eight subpopulations only for $r>2.6$. (B) The average FI of the pinned group tended to be higher than the two neighboring groups for $r>2.6$, although this difference was significant only for a comparatively narrow parameter range. Both these results were contradictory to the observations from the experiments (Fig 1) and the simulations mimicking the experiments (Fig 3), indicating that the effect of pinning at the subpopulation level interacts with the extinction probability
doi:10.1371/journal.pone.0000233.g007

[10]. Different strengths of pinning (Fig 9B) or different rates of migration (Fig 10) did not predict any change in the dynamics, although again there was an overall increase in the magnitude of FI. Considering all these observations together, it is clear that although extinction plays a crucial role in determining subpopulation dynamics, it is not expected to interact with the effects of pinning at the metapopulation level, except when there is variation in the density of pinning.

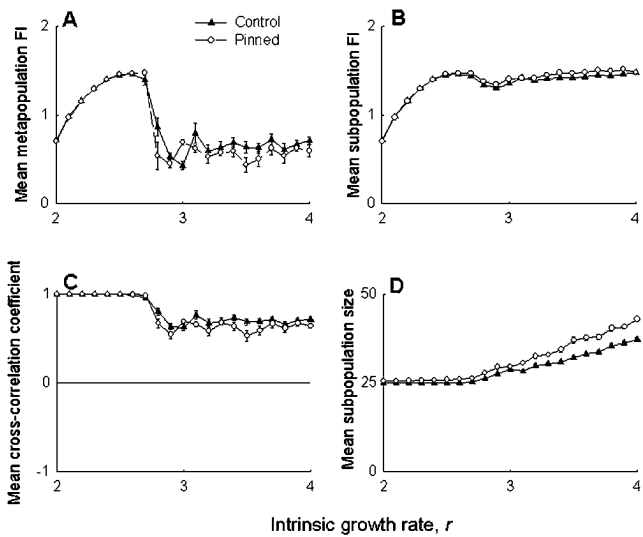

**Figure 8. Simulations with no extinctions: effect of pinning at the metapopulation level.** Although there were qualitative differences in the shapes of the profiles compared to the case when extinction probabilities were incorporated (Fig 4), there were no systematic differences between the control and the pinned treatments in terms of (A) metapopulation FI, (B) subpopulation FI, and (C) subpopulation synchrony. However, the average subpopulation size (D) of the pinned subpopulations was predicted to be similar to the controls for $r<3$, which agrees with the experiments (Fig 2D), but not the earlier simulations (Fig 4D). Taken together it can be said that even in the absence of extinctions, pinning is unlikely to affect metapopulation dynamics.
doi:10.1371/journal.pone.0000233.g008

## Conclusions

Pinning has been suggested as a possible method for stabilizing threatened populations living in a fragmented habitat [9,19]. However, our study indicates that, under a more realistic scenario of noisy, small, extinction-prone subpopulations, constant localized addition of individuals from the outside is not expected to have a major impact on the metapopulation dynamics, and that these results are generalizable. We show that although pinning might interact with extinctions in producing the observed dynamics at the subpopulation level, this is unlikely to affect the metapopulation dynamics. We predict that when there are no local extinctions, increasing the number of pinned subpopulations is likely to destabilize the metapopulation in terms of increased fluctuation in metapopulation size. This result is of potential interest to conservation biologists planning reintroduction of species into natural habitats to boost an extant population, or agricultural scientists trying to eradicate a pest. However, we would like to explicitly point out that our results were derived from simulations based on the Ricker model and it is possible that species whose dynamics are not well approximated by the Ricker might respond differently to pinning [29].

## MATERIALS AND METHODS

### Experimental populations

In this experiment we used eight replicate metapopulations of the fruit fly, *Drosophila melanogaster*, each consisting of nine subpopulations. Four of these metapopulations were subjected to pinning and the other four acted as controls. The seventy-two subpopulations, each represented by a single-vial culture, were derived from a long-standing, outbreeding laboratory population (JB1) of *D. melanogaster*, whose ancestry and maintenance regime has been described elsewhere [31]. Each subpopulation was initiated by placing exactly 20 eggs in a 30 ml glass vial containing ~1 ml of banana-jaggery medium. The flies that came out of these eggs were designated as generation 0, and no direct control was exercised on the egg-density in a vial from that point onwards. Once the adults started eclosing around day 8–9, they were collected daily in corresponding holding vials containing ~3 ml of medium. The adults were transferred to fresh holding vials every alternate day, until day 18 after egg collection. Extreme care was taken to ensure one-to-one correspondence between egg vials and adult collection vials. On day 18, the flies were supplied with excess live-yeast paste for three days, to enhance their fecundity. On the 21st day after collection of eggs, the adult flies were sexed, censused, subjected to migration (see below), and allowed to lay eggs for 24 hours in vials containing ~1 ml banana-jaggery medium. After oviposition, the adults were discarded while the eggs formed the next generation. This maintenance regime (low larval and high adult food levels) has been extensively studied and is known to induce large amplitude periodic oscillations in population numbers [28,32–34].

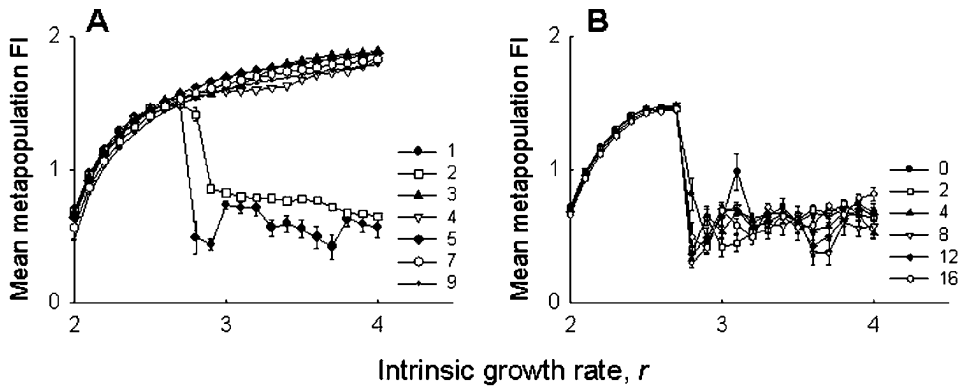

**Figure 9. Simulations with no extinctions: effect of pinning density and magnitude on stability.** (A) When there are no extinctions, increasing the number of pinned patches was generally found to destabilize the metapopulation dynamics, similar to the experimental scenario (5A). (B) Changing the strength of pinning, however, did not affect the metapopulation stability, although there was a change in the FI profile relative to the earlier simulations incorporating extinctions (cf. Fig 5B).
doi:10.1371/journal.pone.0000233.g009

## Migration and pinning

Following an earlier study [21], the subpopulations (single vial *Drosophila* cultures) were arranged on the periphery of a circle, with each of them sending out and receiving migrants to and from the two nearest neighbors. This arrangement can also be visualized as a one-dimensional linear array with periodic boundary condition in terms of migration. Such one-dimensional systems can be found in nature on the shores of lakes or on forest edges. Migration (30%) was imposed by manually removing the required number of flies from a subpopulation and distributing them equally to the two neighboring vials, just prior to reproduction in every generation. Only mated females were migrated, as the dynamics of the population of a sexually reproducing organism is chiefly governed by the number of females. In order to calculate the number of migrant females, the total count in a vial was halved (i.e. a sex ratio of 1:1 is assumed) and rounded upwards in case of fractions. This number was multiplied by 0.3 (i.e. the migration rate) and rounded off to the nearest even integer, to give the total number of female migrants. There were frequent extinctions in the subpopulations during the course of the experiment. Upon extinction, a vial remained empty until it was recolonized by migrants from a neighboring vial.

Pinning was imposed on four metapopulations by introducing eight mated females every generation to a designated (pinned) subpopulation just before the census. The flies required for this purpose were generated from backup vials that had excess (~6 ml) food for larvae and yeast supplement for the adults, and were run in parallel with the experimental vials. It should be noted that for a particular metapopulation, the same subpopulation was pinned in every generation. The average subpopulation size in these experiments was found to be ~26. Thus, the strength of pinning used in this experiment is ~33% of the average population size. Given that only mated females were migrated, this represents a fairly strong perturbation. Since the pinning flies were introduced prior to the census, a 30% migration rate ensured that at least one female was migrated to each of the neighboring vials. Thus, the pinning strength can also be described as six flies to the pinned vial and one each to the two immediate neighbors. The remaining four metapopulations experienced 30% migration but no pinning and, thus, served as controls.

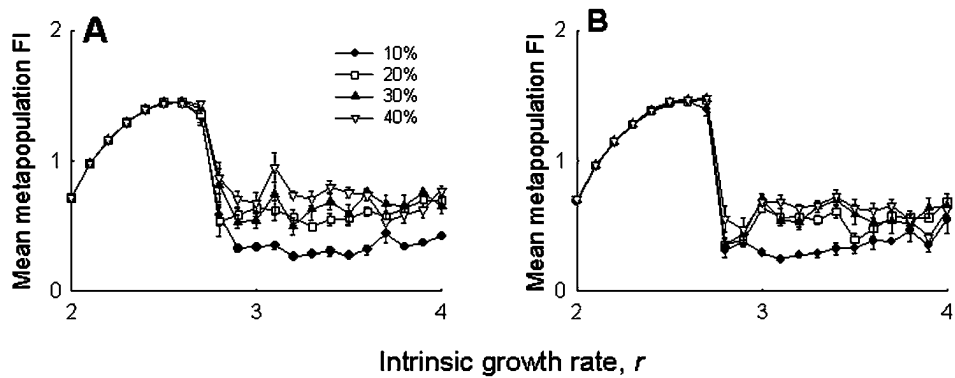

**Figure 10. Simulations with no extinctions: effects of migration rate on stability.** In the absence of extinctions, there were no major differences in the FI of the (A) control and (B) pinned metapopulations. However, there was a change in the profile of the metapopulation FI (cf. Fig 6), indicating that migration rate can interact with the levels of extinction, although this is not expected to interact with pinning to alter the empirically observed patterns of metapopulation stability.
doi:10.1371/journal.pone.0000233.g010

## Measuring stability and synchrony

We considered a population whose size fluctuates with higher amplitude across time to be less stable than one that has lower amplitude of fluctuation ("constancy" attribute of stability, *sensu* [35]). We measured the constancy stability using the fluctuation index (FI) [21], which is given by:

$$FI = \frac{1}{T\bar{N}} \sum_{t=0}^{T-1} abs(N_{t+1} - N_t) ,$$

where $N_t$ is the population size at time $t$ and $\bar{N}$ is the mean population size over $T$ generations. Thus, FI measures the average one-step fluctuation in population size across generations, scaled by the mean population size. Since it is a dimensionless quantity, FI can be used to compare the dynamics of populations even if they vary widely in size. We measured synchrony as the cross-correlation at lag zero of the first differenced time series of log abundance [ln $(N_{t+1}) - \ln(N_t)$, where $N_t$ is the population size at time $t$] of the nearest neighboring subpopulations in a metapopulation [36].

## Statistical analysis

All data were subjected to mixed model analysis of variance (ANOVA), treating replicate metapopulation as a random factor nested within treatment (control/pinned; fixed factor). All statistical analyses were performed using the commercially available software, STATISTICA ® v 5.0 (Statsoft Inc).

## Simulations

The simulation study was designed to be as close to the experimental system as possible. The subpopulation dynamics were modeled using the Ricker model $[n_{t+1} = n_t \exp (r(1 - n_t/K))]$ [23], where $n_t$ represents the subpopulation size at time $t$, and $r$ and $K$ refer to the intrinsic per capita growth rate of the subpopulation and carrying capacity of the patch, respectively. A metapopulation consisted of nine linearly arranged subpopulations, with nearest neighbor migration under periodic boundary condition [21,37]. The carrying capacity, $K$ (25) and the initial subpopulation size (20) were kept constant, unless explicitly stated. All the subpopulations in a given run had the same value of $r$ with a noise term $\varepsilon$ ($0 < \varepsilon < 0.2$; uniform random distribution) added to $r$ for each subpopulation at every generation, to simulate stochastic variation in population growth rates. We simulated the experimental system for $r$ values of the subpopulations ranging from 2.0 to 4.0 in increments of 0.1 and for each value of $r$, we plotted the means and standard errors of 10 independent runs. Estimates of $r$ (mean 2.9; SD .33) and $K$ (mean 25.1; SD 7.2) were derived by fitting the Ricker map to the individual, untransformed population time series from the experimental controls (see *Parameter estimation* below). Thus, the chosen parameter range includes the biologically relevant range for our laboratory populations of *Drosophila*.

Coupled map lattices can have very long transients (supertransients) lasting for thousands of iterations [38], and the behavior of the system during this transient phase can be very different from the equilibrium behavior [39]. Although most theoretical studies on coupled map lattices concentrate on the equilibrium dynamics (eg. [8,40]), we calculated the various metrics estimated in the experiment (see above) using data from only the first 100 iterations, thus concentrating explicitly on the transients. We consider this to be a closer approximation to our experiment, which lasted for 21 generations. Moreover concentrating on transient dynamics is also ecologically more meaningful as any real population is unlikely to experience constant environment, or for that matter even survive, for thousands of generations in nature. (see [41] for a review).

In the simulations seeking to imitate the experimental conditions, the rate of migration was kept constant at 30%. Pinning was modeled by adding 8 individuals to a particular subpopulation (the *pinned* subpopulation), in every generation, prior to migration. Since a Ricker map does not take zero values, we stipulated extinction probabilities that were estimated from the time series of the controls (Fig 11). For this, we calculated the frequency of extinction (absence of at least 1 male and 1 female, before migration) in the next generation $(t+1)$, when the population sizes were low ($<10$), medium ($\geq 10$ and $<70$) or high ($\geq 70$) in the parent generation $(t)$. At an $r$-value of 2.8, this set of extinction probabilities predicted an average of 5.02 out of 9 subpopulations going extinct per generation, which was higher than the corresponding estimation from the experimental controls (3.69). We also computed the extinction probability profile from the experimental data for bin sizes of $<5$, 5–70, and $>70$, and repeated all the simulations with these values of extinction probabilities (data not shown). This predicted an average subpopulation extinction rate of 3.3 out of 9 per generation but did not lead to any qualitatively different predictions at the subpopulation or metapopulation level from those shown in figures 3–6. This suggests that our simulation results are robust to the way in which the extinction probabilities are computed. Varying the initial sizes of the subpopulations (16,18, 20, 22, 24) also failed to affect dynamics (data not shown).

We then studied the effects of pinning different numbers of subpopulations (1, 2, 3, 4, 5, 7, 9), pinning strengths (0, 2, 4, 8, 12, 16 individuals per generation) and migration rates (10%, 20%, 30%, 40%) on the metapopulation dynamics. Since it is known that the distribution of pinned patches can affect the dynamics [9], for a given level of number of pinned patches (i.e. 1, 2, 3, 4, 5, 7 or 9), the spatial arrangement of the pinned patches was kept similar in all simulations. In all the simulations described in this paragraph, the default values of parameters not under investigation were kept constant at levels described for the simulations mimicking the experiments. Thus, for example, in the simulations on the effects of pinning different numbers of subpopulations, the pinning strengths and migration rates were kept constant at 8 individuals and 30%, respectively, and so on.

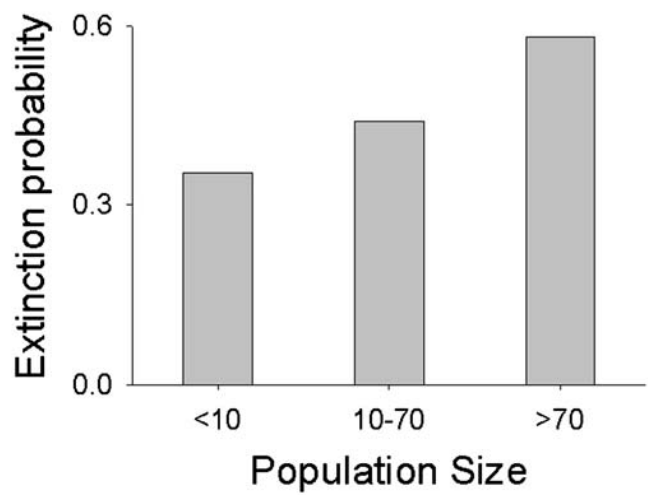

**Figure 11.** Empirically observed extinction probabilities at different population sizes. This shows the fraction of times a population went extinct in generation $t+1$, when the population size in generation $t$ fell within a particular range in the controls.
doi:10.1371/journal.pone.0000233.g011

## Parameter estimation

The least-square estimates of the parameters $r$ and $K$, were obtained using the in-built Quasi-Newton algorithm of STATIS-TICA ® v 5.0 (Statsoft Inc) and, on an average, the model was able to explain ~40% of the variation in the data. While this fraction does appear to be somewhat low, we note that the subpopulations were also undergoing migration in every generation, a fact that was ignored during the modeling procedure, when individual subpopulation time series data were fit to the model. Moreover, the sources of noise in our model are a) white noise in the parameter $r$, and b) experimentally derived extinction probabilities, whereas a model that explicitly incorporates demographic stochasticity [42] might be better suited to model extinction prone populations. While it would be interesting to compare the parameter estimates derived from such detailed models with the estimates obtained in the present study, such an exercise is clearly beyond the scope of the present paper.

## ACKNOWLEDGMENTS

We are grateful to Mallikarjun Shakarad, M. Rajamani and N. Raghavendra for crucial help in fly handling, and Dylan Childs for helpful comments on the manuscript. J. Mohan, Shampa Ghosh, N. M. Satish, N. Rajanna and M. Manjesh also helped in diverse ways during the experiment.

## Author Contributions

Conceived and designed the experiments: AJ SD. Performed the experiments: SD. Analyzed the data: SD. Wrote the paper: AJ SD.